\documentclass[twocolumn,showpacs,footinbib,preprintnumbers]{revtex4}
\usepackage{bm,enumerate,dcolumn,graphicx,color,amsmath,amssymb}
\newcommand{\comment}[1]{}
\newcommand{\bra}{\langle}
\newcommand{\ket}{\rangle}
\newcommand{\eye}{\rm i}

\begin{document}

\title{High-order harmonic generation from Rydberg states at fixed Keldysh parameter}
\author{E. A. Bleda$^1$, I. Yavuz$^2$, Z. Altun$^2$ 
and T. Topcu$^{3,4}$ }
\address{
$^1$Department of Mathematics and Computer Science, Istanbul Arel University, 
34537, Buyukcekmece, Istanbul, TURKEY \\
$^2$Department of Physics, Marmara University, 34722, Ziverbey, Istanbul, TURKEY \\
$^3$Department of Physics, Auburn University, Alabama 36849-5311, USA \\
$^4$Department of Physics, University of Nevada, Reno, NV 89557, USA 
}

\begin{abstract}

Because the commonly adopted viewpoint that the Keldysh parameter $\gamma $ determines 
the dynamical regime in strong field physics has long been demonstrated to be misleading, 
one can ask what happens as relevant physical parameters, such as laser intensity and 
frequency, are varied while $\gamma$ is kept fixed. We present results from our one- and 
fully three-dimensional quantum simulations of high-order harmonic generation (HHG) from 
various bound states of hydrogen with $n$ up to 40, where the laser intensities and the 
frequencies are scaled from those for $n=1$ in order to maintain a fixed Keldysh parameter 
$\gamma$$< 1$ for all $n$. We find that as we increase $n$ while keeping $\gamma $ fixed, 
the position of the cut-off scales in well defined manner. Moreover, a secondary plateau 
forms with a new cut-off, splitting the HHG plateau into two regions. First of these 
sub-plateaus is composed of lower harmonics, and has a higher yield than the second one. 
The latter extends up to the semiclassical $I_p+3.17U_p$ cut-off. We find that this structure 
is universal, and the HHG spectra look the same for all $n\gtrsim 10$ when plotted as a 
function of the scaled harmonic order. 
We investigate the $n$-, $l$- and momentum distributions to elucidate the physical mechanism 
leading to this universal structure.

\end{abstract}

% pacs, the Physics and Astronomy Classification Scheme

\maketitle
\section{Introduction}\label{sec:intro}

High harmonic generation (HHG) is a nonlinear phenomenon in which atoms interacting 
with an intense laser pulse emit photons whose frequencies are integer multiples of 
the driving laser frequency. The emphatic motivation is the generation of spatially 
and temporally coherent bursts of attosecond pulses with high frequencies covering 
a range from vacuum ultraviolet (VUV) to the soft x-ray region~\cite{hentschel}. 
Filtering the high-frequency part of a high-harmonic spectrum allows the syntheses 
of ultrashort, coherent light pulses with energies in the extreme ultraviolet (XUV) 
part of the spectrum. This allows for tracing and controlling electronic processes 
in atoms, as well as coupled vibrational and electronic processes in 
molecules~\cite{worner,itatani}. Some of the most visible applications of ultrashort 
pulses of attosecond duration involve resolving the electronic structure with high 
degree of spatial and temporal resolution~\cite{chen}, controlling the dynamics in 
the XUV-pumped excited molecules~\cite{rini}, and exciting and probing inner-shell 
electron dynamics with high resolution~\cite{sandberg}. Time-resolved 
holography~\cite{tobey}, imaging of molecular orbitals~\cite{itatani}, and attosecond 
streaking~\cite{itatani2} are also among the state-of-the-art applications of HHG. 

High-order harmonic generation is a process well described within the semi-classical 
three step model (ionization, propagation followed by recombination). The plateau 
region, where consecutive harmonics have approximately the same intensity, constitutes 
the main body of a high-harmonic spectrum. First step of the three step model is the 
tunneling of the electron through the Coulomb potential barrier suppressed by the laser 
field. The second step is laser-driven propagation of the free electron, and the third 
step is the rescattering of the electron with its parent ion.  During this last step, 
the electron can recombine with its parent ion and liberate its excess energy as a 
short wavelength harmonic photon. The three step model predicts that the highest kinetic 
energy that an electron gains during its laser-driven excursion is given by $3.17 U_p$, 
where $U_p = F^2/(4\omega_0^2)$ is the quiver energy of the free electron in the laser 
field, and $F$ and $\omega_0$ are the laser field amplitude and frequency. The highest 
harmonic frequency, $\omega_c$, that can be generated within this 
model is $q_{\max}\omega_0 = \left|{E_b}\right| + 3.17 U_p$, where $\left|{E_b}\right|$ 
is the binding energy of electron in the atom and $q_{\max}$ is the order of the 
cut-off harmonic~\cite{corkum}. 

A crucial assumption in this physical picture is that the electron tunnels into the 
continuum in the first step in a laser field characterized by a small Keldysh parameter. 
This liberates the electron with no excess kinetic 
energy, and its subsequent excursion is driven by the classical laser field alone. 
Keldysh parameter $\gamma$ is commonly used to refer to one of the two dominant ionization 
dynamics in strong fields; tunneling or multiphoton regimes~\cite{keldysh}. 
It is defined as the the time it takes for the electron to tunnel the 
barrier in units of the laser period, i.e., $\gamma \sim \tau/T$. Here 
$\tau$ is the tunneling time and $T=2\pi /\omega_0$ is the laser period. If the tunneling 
time is much smaller than the laser period, one could expect that it is likely for 
the electron to tunnel through the barrier. In contrast, if tunneling time is much 
longer than the laser period, then the electron doesn't have enough time tunnel through 
the depressed Coulomb barrier, and ionization can only occur through photon absorption. 
The Keldysh parameter can be expressed as 
$\gamma =\omega_0\sqrt{2\left|{E_b}\right|}/F$~\cite{keldysh}. 

Although the Keldysh parameter is widely used to refer to the underlying dynamics 
in strong field ionization, there are studies which suggest that it is an inadequate 
parameter in making this assessment~\cite{reiss1,reiss2,topcu12} when a large range of laser 
frequencies are considered. Thus, it is natural to ask what happens in the strong field 
ionization step of HHG as a function of $n$, as relevant parameters, such as laser 
intensity and frequency, are varied while $\gamma$ is kept fixed. In this paper, we 
investigate the HHG process from the ground and the Rydberg states of a hydrogen atom 
using a one-dimensional $s$-wave model supported by fully three-dimensional quantum 
simulations. The central idea is that in a hydrogen atom, both the field strength $F$ 
and the frequency $\omega_0$ scale in a particular fashion with the principal quantum 
number $n$. Scaling the field strength by $1/n^4$ and the frequency by $1/n^3$, it is 
evident that $\gamma =\omega_0\sqrt{2\left|{E_b}\right|}/F$ remains unaffected as $n$ is 
changed, provided that both $F$ and $\omega_0$ are scaled accordingly while $n$ is 
varied. 

In the spirit of the Keldysh theory, going beyond the ground state and starting 
from higher $n$ as the initial state, scaling $F$ and $\omega_0$ to maintain a fixed value 
of $\gamma$ should keep the ionization step of the harmonic generation in the same 
dynamical regime. We calculate HHG spectra starting from the ground state of hydrogen 
using laser parameters for which $\gamma<1$ (tunneling), and then calculate the 
high-harmonic spectra from increasingly larger $n$-states, scaling $F$ and $\omega_0$ 
from the ground state simulations and keep $\gamma$ fixed. If the Keldysh parameter 
is indeed adequate in referring to the ionization step properly in HHG, one should 
expect that the physics of the three-step process would remain unchanged, as the 
remainder of the steps involve only classical propagation of the electron in the 
continuum, and the final recombination step, which is governed by the conservation of 
energy. 

There are a number of studies devoted to HHG from Rydberg atoms. The main motivation 
in these efforts are primarily increasing the conversion efficiency in the harmonic 
generation to obtain higher yields, which in turn would enable the generation of more 
intense attosecond pulses. Hu {\it et al.}~\cite{hu} demonstrated that, by stabilization 
of excited outer electron of the Rydberg atom in an intense field, a highly efficient 
harmonic spectrum could be generated from the more strongly bound inner electrons. 
In another recent study, Zhai {\it et al.}~\cite{zhai1,zhai2} proposed that an enhanced 
harmonic spectrum is possible if the initial state is prepared as a superposition of 
the ground and the first excited state. The idea behind this method is that when coupled 
with the ground state, ionization can occur out of the excited state, initiating the harmonic 
generation. Since the excited state has lower ionization potential than the ground 
state, this in principle can result in higher conversion efficiency if the electron 
subsequently recombines into the excited state. In this scenario, the high-harmonic 
plateau would still cut-off at the semiclassical limit $I_p+3.17U_p$ with $I_p$ being 
that of the excited state. If, however, upon ionization out of the excited state, the 
electron recombines into the ground state, the cut-off can be pushed up to higher 
harmonics. Same principle is also at play in numerous studies proposing two-color 
driving schemes for HHG, with one frequency component serving to excite the ground 
state up to an excited level with a lower ionization potential, thus increasing the 
ionization yield (see for example~\cite{zhai3}). 

In this paper, we report HHG spectra from ground and various Rydberg states with
 $n$ up to 40 for hydrogen atom, where the laser intensity and 
the frequency are such that the ionization step occurs predominantly in the 
tunneling regime. Starting with $\gamma=0.755$ at $n=1$, we go up in $n$ of the initial 
state and scale $F$ by $1/n^4$ and $\omega_0$ by $1/n^3$, keeping $\gamma$ constant. 
We discuss the underlying mechanism in terms of field ionization and final 
$n$-distributions after the laser pulse. We find that the harmonic order of the cut-off 
predicted by the semiclassical three-step model scales as 
$1/n$ when $F$ and $\omega_0$ are scaled as described above, and $\gamma$ is kept fixed. 
We repeat some of these model simulations by solving the fully three-dimensional time-dependent 
Schr\"odinger equation to investigate the effects which may arise due to angular momenta 
in high-$n$ manifolds. For select initial $n$ states, we look at momentum distributions 
of the ionized electrons, and the wave function extending beyond the 
peak of the depressed Coulomb potential at $1/\sqrt{F}$. Unless otherwise stated, we use 
atomic units throughout. 

\section{One-dimensional calculations}\label{sec:theory_hhg1d}

The time-dependent Schr\"{o}dinger equation of an electron interacting with the 
proton and the laser field $F(t)$ in the $s$-wave model in length gauge reads 

\begin{eqnarray}\label{schro}
\eye\; \frac{\partial \psi (r,t)}{\partial t} 
	= \left[ -\frac{1}{2}\frac{{{d}^{2}}}{d{{r}^{2}}}-\frac{1}{r}+rF(t) \right] 
		\psi (r,t).	
\end{eqnarray} 
In our simulations, time runs from $-t_f$ to $t_f$. This choice of time range 
centers the carrier envelope of the laser at $t=0$, which simplifies its mathematical 
expression. We choose the time-dependence of the electric field $F(t)$ to be 

\begin{eqnarray}\label{laser} 
F(t)={F_0}\exp (-(4\ln 2){{t}^{2}}/{{\tau }^{2}})\cos ({{\omega }_{0}}t),	
\end{eqnarray} 
where $F_0$ is the peak field strength, $\omega_0$ is the laser frequency and $\tau $ 
is the field duration at FWHM. Our one dimensional model is an $s$-wave 
model and restricted to the half space $r\ge 0$ with a hard wall at $r=0$. Having a hard 
wall at $r=0$ when there is no angular momentum can potentially be problematic, because 
the electron can absorb energy from the hard wall when using $-1/r$ potential. However, we 
believe that this model is adequate for the problem at hand, because we are deep in the 
tunneling regime. In our calculations, the number of photons required for ionization to occur 
through photon absroption is $\sim$9 for $n=1$, approaches to 71 by $n=10$ and stays so for 
higher $n$. As a result, ionization takes place primarily in the tunneling regime. If an extra 
photon is absorbed at the hard wall, its effect would mostly concern the lowest harmonics, 
which we are not interested in. In Sec.~\ref{sec:theory_hhg3d}, we show that the results 
we obtain in this section are consistent with our findings from fully three-dimensional 
calculations. 

We consider cases in which the electron is initially prepared in an $ns$ state, where $n$ 
ranges from 1 up to 40. Our pulse duration is 4-cycles at FWHM for each case, and the wavelength 
of the laser field is 800 nm for the ground state. This gives a 2.7 fs optical cycle 
when the wavelength is 800 nm. Thus, the total pulse duration $\tau$ for the ground state is 
$\sim$11 fs and it scales as $n^{3}$. For the 4s state, this results in a pulse duration 
of $\sim$704 fs, while it amounts to $\sim$5.6 ps for the 8s state. 

For the numerical solution of equation~(\ref{schro}), we perform a series of calculations 
to make sure that the mesh and box size of the radial grid and the time step we use are 
fine enough so that our results are converged to within a few per cent. As we go beyond the 
1s state, we increase the radial box size to accommodate the growing size of the initial 
state and the interaction region. We propagate Eq.~\eqref{schro} for 
excited states using a square-root mesh of the form ${{j}^{2}}\delta r$, where $j$ is the 
index of a radial grid point, $\delta r=R/{{N}^{2}}$, $R$ is the box size, and $N$ is the 
number of grid points. This type of grid is more efficient than using a uniform mesh in 
problems involving Rydberg states~\cite{topcu07}, because it puts roughly the same number 
of points between the successive nodes of a Rydberg state. For the ground state, the box 
size is $R=750$ a.u. and ${N}=800$, which gives ${\delta r}=0.0012$ a.u.. For excited states, 
the box size grows $\sim{{n}^{2}}$ and with a proper selection of ${\delta r}$, we make 
sure that the dispersion relation $k\delta r=0.5$ holds for each $n$ state, where $k$ is 
the maximum electron momentum acquired from the laser field: 
$k=\sqrt{2{{E}_{\max }}}$ and ${{E}_{\max }}=3.17{{U}_{p}}$. 

The time propagation of the wave-function is carried out using an implicit scheme. For the 
temporal grid spacing $\delta t$, we use $n^3$/180 of a Rydberg period, which is small 
enough to give converged results. A smooth mask function which varies from 1 to 0 starting 
from 2/3 of the way between the origin and the box boundary is multiplied with the solution of 
equation~(\ref{schro}) at each time step to avoid spurious reflections from the box boundaries. 

The time-dependent solutions of equation~(\ref{schro}) are obtained for each initial $ns$ 
state, which we then use to calculate the time-dependent dipole acceleration, 
$a(t)=\bra \ddot{r} \ket (t)$: 
\begin{eqnarray}\label{inten} 
a(t) &=& \bra \psi(r,t) | \left[H, \left[H,r \right] \right] | \psi(r,t) \ket \;. 
\end{eqnarray} 
Because the harmonic power spectrum is proportional to the Fourier transform of the squared 
dipole acceleration, we report $|a(\omega)|^2$ for harmonic spectra. 

The initial wave function is normalized to unity, and the time-dependent ionization 
probability is calculated as the remaining norm inside the spatial box 
at a given time $t$, 
\begin{eqnarray}\label{ion} 
P(t)=1-\int\limits_{0}^{{R}}{{{\left| \psi (r,t) \right|}^{2}}dr}. 
\end{eqnarray} 
In evaluation of the ionization probability, we propagate the wavefunction long enough after 
the pulse is turned off until $P(t)$ converges to a time-independent value. 

%---------------------------------------------------------------------------------
\subsection{Results and discussion}\label{sec:results_hhg1d}

In our one-dimensional simulations, we consider cases where the atom is initially 
in an $n$s state with $n$ up to 40. The laser parameters are critically chosen 
so that the Keldysh parameter is fixed at $\gamma =0.755$ for each initial $n$, 
and the scaled frequency of the laser field is ${{\omega }_{0}}{{n}^{3}}\ll 1$, 
{\it i.e.}, the electric field has a slowly varying time-dependence compared with 
the Kepler period ${{T}_{K}}=2\pi {n}^{3}$ of the Rydberg electron. For example, 
for an 800 nm laser, an optical cycle is $\sim$18 times the Kepler period for $n=1$. 
The cut-off frequency ${{\omega }_{c}}$ predicted by the three-step model is 
${{\omega }_{c}}=\left| {{E}_{b}} \right|+3.17{{U}_{p}}$~\cite{corkum}, where 
${{U}_{p}}=F^{2}/4\omega _{0}^{2}$ is the ponderomotive potential. Since 
$\left| {{E}_{b}} \right|$, ${F}$ and ${{\omega }_{0}}$  scale as ${{n}^{-2}}$, 
${{n}^{-4}}$ and ${{n}^{-3}}$ respectively, the cut-off frequency ${{\omega }_{c}}$ 
scales as ${{n}^{-2}}$ and the harmonic order of the cut-off 
${{q}_{\max }}={{\omega }_{c}}/{{\omega }_{0}}$ scales as $n$ for fixed $\gamma$. 

\begin{figure}[ht!]
  \begin{center}
    \resizebox{160mm}{!}{\includegraphics{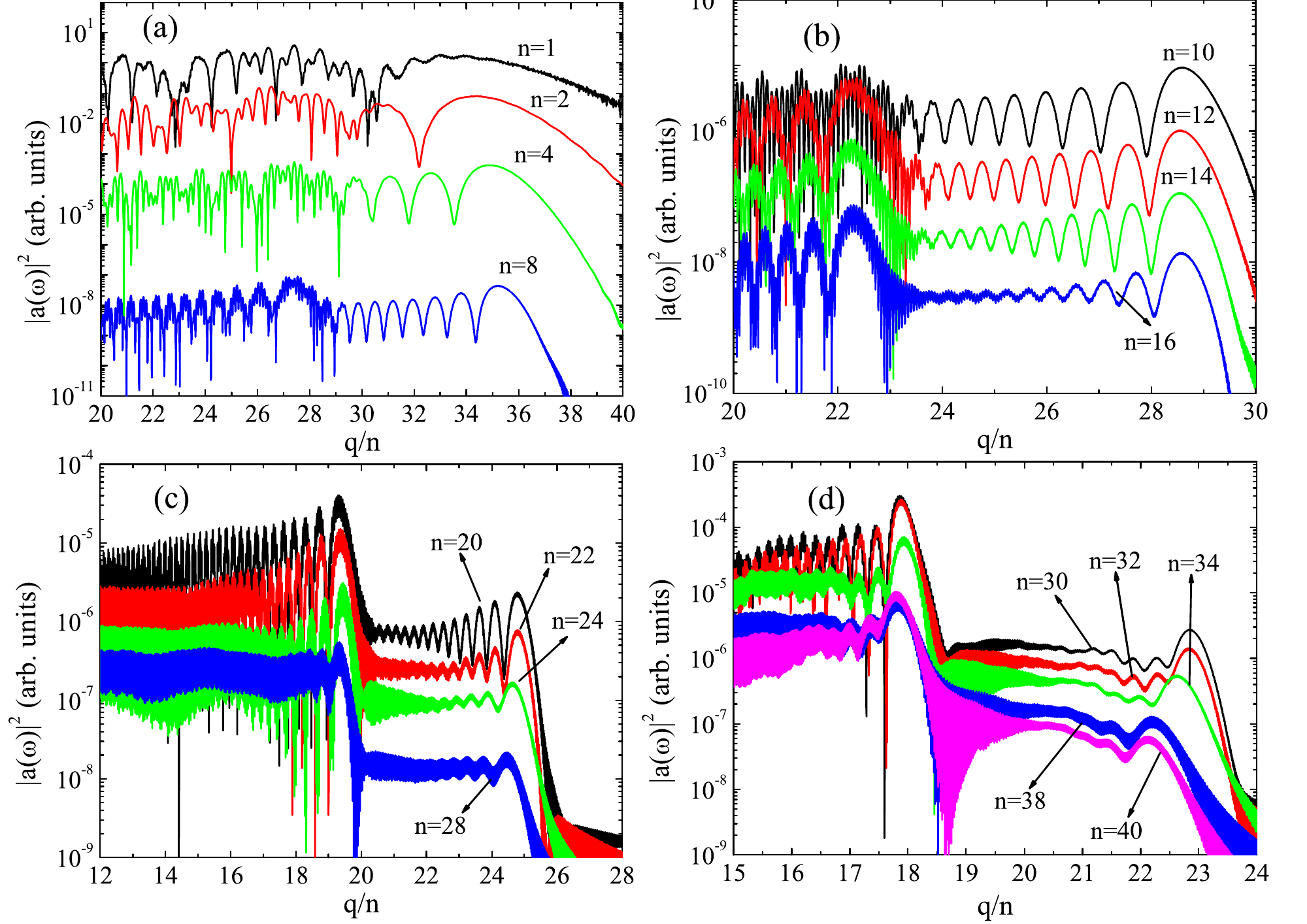}}
  \end{center}
  \caption{(Color online) High harmonic spectrum from the Rydberg states of H atom. 
  	The scaled laser field intensities and the wavelengths are, 
  	(a) 200/${{n}^{8}}$ TW/cm$^2$ and 800${{n}^{3}}$ nm, 
  	(b) 300/${{n}^{8}}$ TW/cm$^2$ and 652${{n}^{3}}$ nm, 
  	(c) 400/${{n}^{8}}$ TW/cm$^2$ and 566${{n}^{3}}$ nm, 
  	(d) 470/${{n}^{8}}$ TW/cm$^2$ and 522${{n}^{3}}$ nm. 
  	The width of the laser pulse is 4-cycles at FWHM, and the selected parameters 
  	correspond to $\gamma =0.755$ in each case. The scaled harmonic order is 
  	$q/n$, where $q=\omega /{{\omega }_{0}}$ is the harmonic order. 
  }
  \label{fig_hhg1d}
\end{figure}

Harmonic spectra from these simulations are seen in Fig.~\ref{fig_hhg1d} (a)-(d) as 
a function of the scaled harmonic order $\widetilde{q}=q/n$, where $q=\omega/\omega_0$ 
is the harmonic order. In Fig.~\ref{fig_hhg1d} (a), the scaled laser intensity and 
the wavelength are $200/{{n}^{8}}$ TW/cm$^2$ and 800${{n}^{3}}$ nm, which correspond to 
$\gamma =0.755$. The most prominent feature in these spectra is a clear double plateau 
structure, exhibiting one plateau with a higher yield and another following with lower 
yield. The second plateau terminates at the usual semiclassical cut-off. These plateaus 
are connected with a secondary cut-off, which converges to a fixed scaled harmonic order 
$\widetilde{q}=q/n$ as $n$ becomes large. 

We also note that the overall size of $|a(\omega)|^2$ drops significantly with increasing 
$n$ in Fig.~\ref{fig_hhg1d} (a). For example, going from $n=2$ to $n=4$, $|a(\omega)|^2$ 
drops about 3 orders of magnitude, and from $n=4$ to $n=8$ it drops roughly 4 orders of 
magnitude. The spectrum obtained for $n=8$ is about 9 orders of magnitude lower than that 
for $n=1$. Beyond $n=8$, the overall sizes of the spectra are too small and plagued by 
numerical errors, which is why we stop at $n=8$ in panel (a). This is because the amplitude 
of the wave function component contributing to the three-step process is too small to yield 
a meaningful spectrum within our numerical precision. 
In order to ensure sizable HHG spectra while climbing up higher in $n$, we adopt the 
following procedure: We split the Rydberg series into different groups of initial 
$n$-states, which are subject to different laser parameters but have the same $\gamma$ 
value within themselves. Within each group, we climb up in $n$ by scaling the laser 
parameters for the lowest $n$ in the group until $|a(\omega)|^2$ becomes too small. 
We then move onto the next group of $n$-states, increasing the laser intensity and 
the frequency ($\gamma \propto \omega/F$) for the lowest $n$ in the group 
while attaining the same $\gamma$ as in the previous $n$-groups. Scaling this intensity 
and frequency, we continue to climb up in $n$ until again $|a(\omega)|^2$ becomes too 
small, at which point we terminate the group and move onto the next. 

Following this procedure, we are able to achieve HHG spectra for states up to $n=40$ in 
Fig.~\ref{fig_hhg1d}. The first $n$-group in panel (a) includes states between $n=1-8$, 
and the laser intensity and wavelength are $200/{{n}^{8}}$ TW/cm$^2$ and 800${{n}^{3}}$ nm. 
In panel (b) is the second group with  $n=10-18$ and the laser parameters $300/{{n}^{8}}$ 
TW/cm$^2$ and 652${{n}^{3}}$ nm. In panel (c), $n=20-28$ and the laser parameters are 
$400/{{n}^{8}}$ TW/cm$^2$ and 566${{n}^{3}}$ nm, and finally in panel (d), $n=30-40$ with 
intensity and wavelength $470/{{n}^{8}}$ TW/cm$^2$ and 522${{n}^{3}}$ nm. 
The peak field strengths corresponding to these intensities are lower than the critical field 
strengths for above-the-barrier ionization for the states we consider, and the ionization 
predominantly takes place in the tunneling regime.

\begin{figure}[h!tb]
	\begin{center}$
		\begin{array}{c}
		   \resizebox{100mm}{!}{\includegraphics{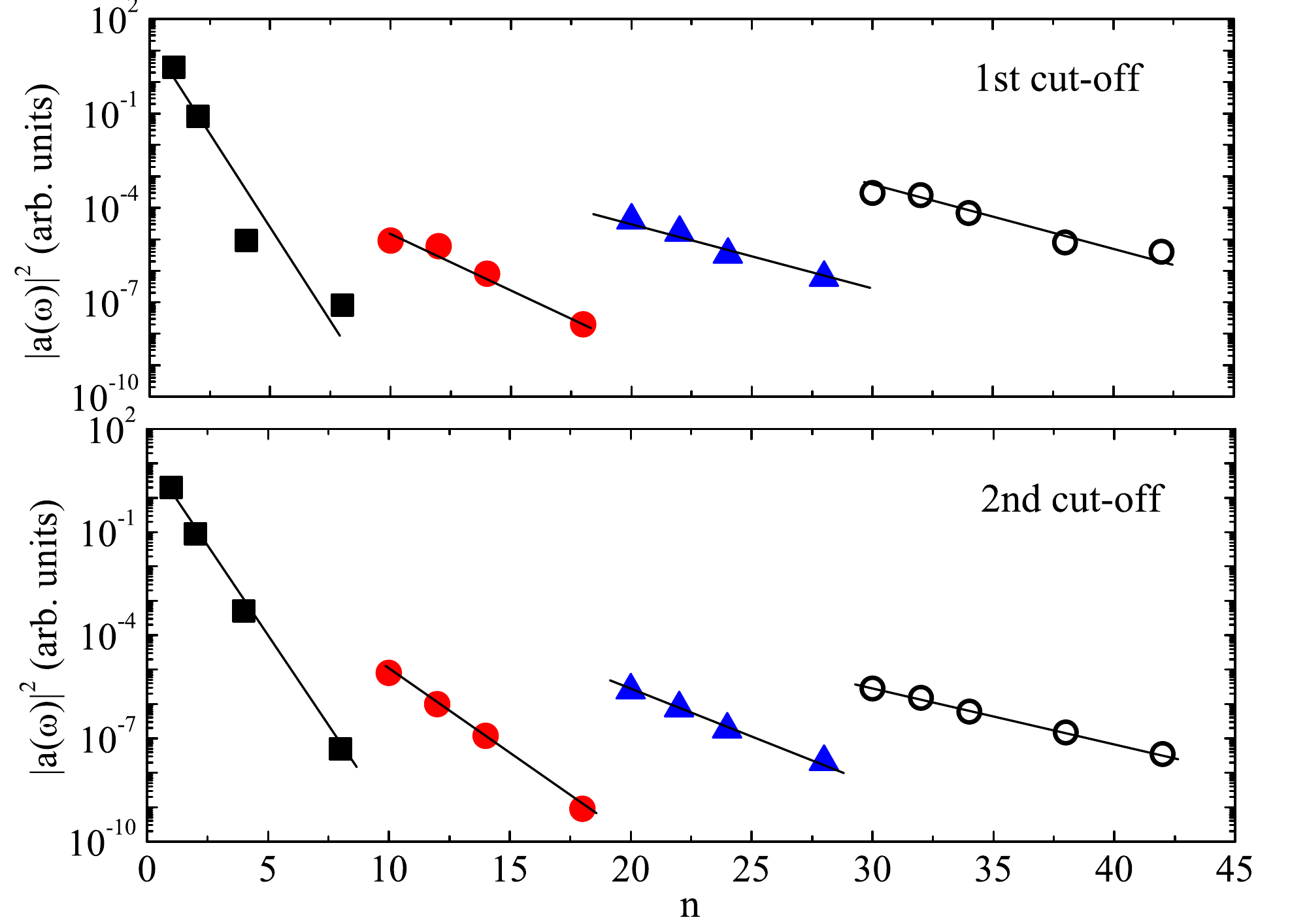}} \\
       \resizebox{103mm}{!}{\includegraphics{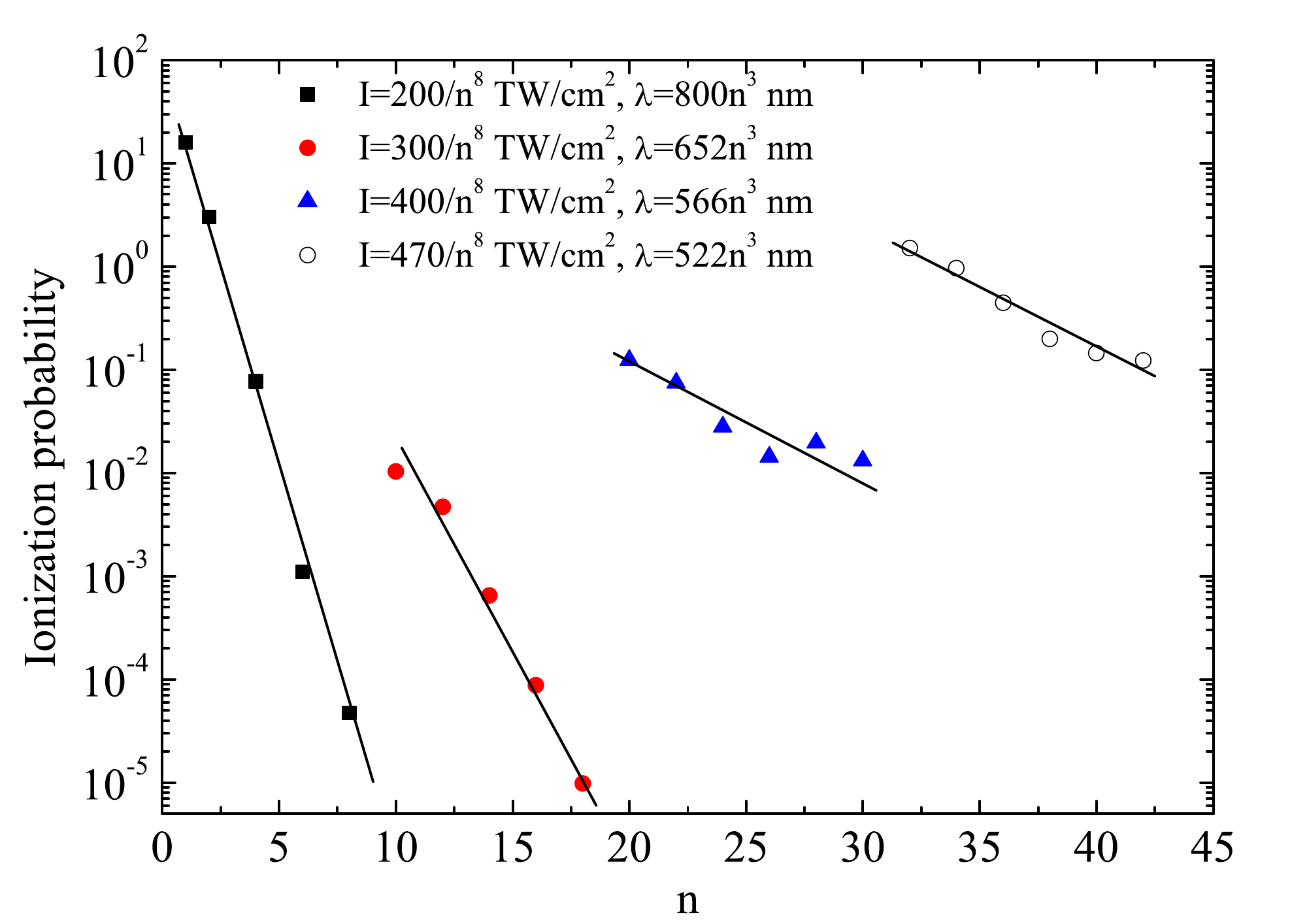}}  
         \end{array}$
     \end{center}
  \caption{(Color online) (Upper two panels) $|a(\omega)|^2$ at the 1st and 2nd cut-offs 
  	of the H atom as a function of $n$, obtained from Fig~\ref{fig_hhg1d} (a)-(d). 
  	Within each $n$-group, the $|a(\omega)|^2$ drops with increasing $n$. 
  	(Lower panel) Ionization probabilities of H atom as a function 
  	of $n$ mimic the behavior of $|a(\omega)|^2$ in the upper panels. 
  	The field parameters are the same as in Fig.~\ref{fig_hhg1d} (a)-(d). 
  }
  \label{fig_hhg1d_iprob}
\end{figure}

% Fig. 02 upper two panels 
The dipole accelerations at the two cut-off harmonics for each $n$-group seen in 
Fig.~\ref{fig_hhg1d} (a)-(d) are plotted in the upper two panels of Fig.~\ref{fig_hhg1d_iprob}. 
Here, we plot $|a(\omega)|^2$ as a function of $n$. This figure points to a situation in 
which $|a(\omega)|^2$ drops with increasing $n$ within each group of $n$. Also, for the first 
few $n$-groups, $|a(\omega)|^2$ drops much faster compared to those involving higher $n$. 
%Furthermore, we find that within each $n$-group, $|a(\omega)|^2$ at the first and the second 
%cut-offs appear to scale almost linearly. 
The reason for the decreasing $|a(\omega)|^2$ within each $n$-group in 
Fig.~\ref{fig_hhg1d_iprob}, can be understood by calculating the ionization probabilities in 
each case, and examining how it changes as $n$ is varied. 

% Fig. 02 lower panel 
Although completely ionized electrons do not contribute to the HHG process, ionization and 
HHG are two competing processes in the tunneling regime. As a result, decrease in 
one alludes to decrease in the other. The ionization probabilities from the $n$s states 
in Fig.~\ref{fig_hhg1d} are plotted against their principal quantum numbers in the lowest 
panel of Fig.~\ref{fig_hhg1d_iprob}. It is clear that as we go beyond the ground state, the 
ionization probabilities drop significantly as $n$ is increased within each group. This 
decrease is rather sharp for the first group and it levels off as we go to successive groups 
involving higher $n$. The values of the scaled frequencies $\Omega={\omega}{n}^{3}$ are the 
same in each $n$-group, and the laser parameters are chosen so as to make sure the condition 
$\Omega \ll 1$ holds. This ensures that the ionization is not hindered by processes such as 
dynamic localization. The reason behind the decreasing ionization probabilities within each 
$n$-group can be understood using the quasiclassical formula~\cite{keldysh} for the tunneling 
ionization rate: 
\begin{eqnarray}\label{keldysh}
{{\Gamma }_{K}}\propto {{\left( \left| {{E}_{b}} 
\right|{{F}^{2}} \right)}^{1/4}}
\exp \left( -2{{(2\left| {{E}_{b}} \right|)}^{3/2}}/3F \right) \; .
\end{eqnarray} 
The laser field intensity and electron binding energy scale as $\sim$$1/{{n}^{4}}$ and 
$\sim$$1/{{n}^{2}}$. Thus, the exponent in the exponential factor in $\Gamma_K$ scales 
as $1/n$, which results in decreasing ionization probabilities within each $n$-group when 
plotted as a function of $n$ in the lowest panel of Fig.~\ref{fig_hhg1d_iprob}. 
This behavior is reflected in the corresponding HHG spectra in Fig.~\ref{fig_hhg1d} 
and the upper panels in Fig.~\ref{fig_hhg1d_iprob} as diminishing of the HHG yield.

The decrease in the ionization probability also slows down as as we successively move onto 
groups of higher $n$, as indicated by the decreasing slopes of the ionization probabilities 
in Fig.~\ref{fig_hhg1d_iprob} between successive $n$-groups. We find that the ratio of the 
ionization probabilities between the 2s and 4s states in Fig.~\ref{fig_hhg1d_iprob} is 
$\sim$39, whereas between the 12s and 14s states it is $\sim$7, between 22s and 24s states 
$\sim$3, and between 32s and 34s states $\sim$2. This is an artifact of the scheme we employ 
in which we divide up the Rydberg series into successive groups of $n$s states to ensure 
sizable HHG spectra. The rate of decrease in the ionization probability in each group is 
determined by the slope of $\Gamma_K$, {\it i.e.}, ${\rm d}\Gamma_K /{\rm d}n$. This slope 
is proportional to the laser intensity we pick for the lowest $n$ in each group in order to 
initiate it, and we scale it down by $1/n^8$ inside the group to keep $\gamma$ fixed. 
However, although this start-up intensity for each group is larger than what it would have 
been of we were to continue up in $n$ in the previous group, it is still smaller than the 
initial intensity in the previous group. This results in a decreased slope going through 
successive $n$-groups. Hence the decay rates for the ionization probability in successive 
groups taper off, which is reflected in the two upper panels in Fi.g~\ref{fig_hhg1d_iprob}.

\begin{figure}[ht!]
  \begin{center}
    \resizebox{160mm}{!}{\includegraphics{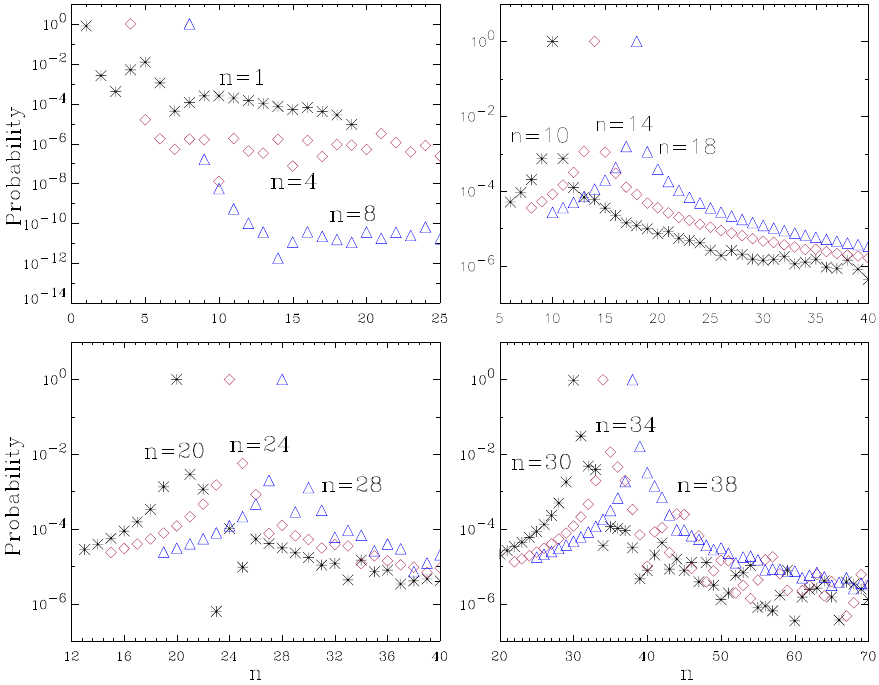}}
  \end{center}
  \caption{(Color online) The probability distributions in $n$ following the laser pulse 
  for the initial states seen in Fig.~\ref{fig_hhg1d}. It is clear that the atom essentially 
  resides in its initial state after the pulse, which means the recombination step 
  in the harmonic generation process occurs primarily back to the initial state. 
  The probability to find the atom in other nearby states is orders of magnitude 
  smaller, and the probability distribution becomes symmetrical about the initial 
  state for $n>10$ due to decreasing anharmonicity in the surrounding energy level 
  structure. 
  }
  \label{fig_hhg1d_ndist}
\end{figure}

% Fig. 3 final n-dependence
We also calculate the final $n$-distributions for the atom after the laser pulse to see the 
extent of $n$ mixing which may have occurred during its evolution in the laser field. This is 
done by allowing the wave functions to evolve according to Eq.~\eqref{schro} long enough 
after the laser pulse to attain a steady state. We then project them onto the bound eigenstates 
of the atom to determine the final probability distributions $P(n)$ to find the atom in a 
given bound state. The results are shown in Fig.~\ref{fig_hhg1d_ndist}. It is evident from 
the figure that most of the wavefunction resides in the initial state after the laser pulse, 
and that there is small amount of mixing into adjacent $n$ states. The mixing is small 
because only a small fraction of the total wavefunction takes part in the HHG process. 
However, we cannot deduce from our calculations what fraction of the wavefunction actually  
participates in HHG, and hence what fraction of it spreads to higher $n$. 
Because the HHG and ionization are competing processes in this regime, the ionization 
probabilities seen in the bottom panel of Fig.~\ref{fig_hhg1d_iprob} can be taken to be 
an indication of the amplitude that goes into the HHG process. For example, 
at $n=4$, the ionization probability is at $\sim$10\% level in Fig.~\ref{fig_hhg1d_iprob} 
and the largest amplitude after the laser pulse is in $n=5$ in Fig.~\ref{fig_hhg1d_ndist} 
at 10$^{-5}$ level. This indicates that roughly a part in $10^6$ of the amplitude participating 
in the HHG process recombines into higher $n$-states. On the other hand, at $n=20$, 
the ionization probability is also at $\sim$10\% level, but the spreading in $n$ is 
between $\sim$1\% and $\sim$0.1\% level, suggesting that between roughly 1 and 10\% of the 
wavefunction participating the HHG process gets spread over adjacent $n$. 
In the recombination step of the HHG process, the probability for recombination back into the 
initial state is the largest, chiefly because the electron leaves the atom through 
tunneling with no excess kinetic kinetic energy. It largely retains the character of the 
initial state because its subsequent excursion in the laser field is classical and mainly 
serves for the electron wavepacket to acquire kinetic energy before recombination. In the 
next section, we discuss how this small spread helps shape the double plateau structure 
seen in Fig.~\ref{fig_hhg1d}.

\section{Three-dimensional Calculations} \label{sec:theory_hhg3d}

Three dimensional quantum calculations were carried out by solving the 
time-dependent Schr\"odinger equation as described in Ref.~\cite{topcu07}. For 
sake of completeness, we briefly outline the theoretical approach below. We 
decompose the time-dependent wave function in spherical harmonics 
$Y_{\ell,m}(\theta,\phi)$ as 
\begin{equation}\label{scheq3d:decomp}
\Psi(\vec{r},t)=\sum_\ell f_\ell(r,t) Y_{\ell,m}(\theta,\phi)
\end{equation}
such that the time-dependence is captured in the coefficient $f_\ell(r,t)$. 
For each angular momenta, $f_\ell(r,t)$ is radially represented on a square-root 
mesh, which becomes a constant-phase mesh at large distances. This is ideal for 
description of Rydberg states on a radial grid since it places roughly the same 
number of radial points between the nodes of high-$n$ states. On a square-root 
mesh, with a radial extent $R$ over $N$ points, the radial coordinate of points 
are $r_j=j^2\delta r$, where $\delta r=R/N^2$. 
We regularly perform convergence checks on the number of angular momenta we 
need to include in our calculations as we change relevant physical parameters, 
such as the laser intensity. 
For example, $\delta r=4\times10^{-4}$ a.u. in a $R=2000$ a.u. box 
gave us converged results for $n=4$, whereas $\delta r=8\times10^{-4}$ a.u. in a $R=2800$ a.u. 
box was sufficient at $n=8$. We also found that the number of angular momenta we needed 
to converge the harmonic spectra was much larger than $n-1$ for an initial $n$ state 
({\it e.g.}, $\sim$120 for the $n=8$ state). 

We split the total hamiltonian into an atomic hamiltonian plus the interaction 
hamiltonian, such that $H(r,l,t)=H_\text{A}(r,l) + H_\text{L}(r,t)-E_\text{0}$. Note 
that we subtract the energy of the initial state from the total hamiltonian to 
reduce the phase errors that accumulate over time. The atomic hamiltonian 
$H_\text{A}$ and the hamiltonian describing the interaction of the atom with the 
laser field in the length gauge are 
\begin{eqnarray}\label{scheq3d:split}
H_\text{A}(r,l) &=& -\frac{1}{2}\frac{\text{d}^2}{\text{d}r^2}-\frac{1}{r}
	+\frac{l(l+1)}{2r^2} \;, \\
H_\text{L}(r,t) &=& F(t) z \cos(\omega t) \;.
\end{eqnarray}

Contribution of each of these pieces to the time-evolution of the wave function 
is accounted through the lowest order split operator technique. In this
technique, each split piece is propagated using  an implicit scheme of order 
$\delta t^3$. A detailed account of the implicit method and the split operator 
technique employed is given in Ref.~\cite{topcu07}. 
The interaction Hamiltonian, $F(t)r\cos(\theta)$, couples $\ell$ to 
$\ell \pm 1$. The laser pulse envelope has the same time-dependence as in the 
one-dimensional $s$-wave model calculations (Eq.~\ref{laser}). 

The harmonic spectrum is usually described as the squared Fourier transform of the 
expectation value of the dipole moment ($d_z(t)=\bra z\ket(t)$), dipole velocity 
($v_z(t)=\bra \dot{z}\ket(t)$), or the dipole acceleration ($a_z(t)=\bra \ddot{z}\ket(t)$) 
(see~\cite{bandrauk} and references therein). 
In our three-dimensional calculations, we evaluate all three forms and compare 
them for different initial $n$ states: 
\begin{eqnarray}
d_z(t) &=& \bra \Psi(\vec{r},t) |z| \Psi(\vec{r},t) \ket \\
v_z(t) &=& \bra \Psi(\vec{r},t) |\dot{z}| \Psi(\vec{r},t) \ket \\
a_z(t) &=& \bra \Psi(\vec{r},t) |\ddot{z}| \Psi(\vec{r},t) \ket  \;,
\end{eqnarray} 
where $\dot{z}=-\eye[H,z]$ and $\ddot{z}=-[H,[H,z]]$. 
Ref.~\cite{bandrauk} found that the Fourier transforms $d_z(\omega)$, $v_z(\omega)$, 
and $a_z(\omega)$ 
are in good agreement when the pulses are long and ``weak" in harmonic generation 
from the ground state of H atom, where ``weak" refers to intensities below over-the-barrier 
ionization limit. As we increase the initial $n$ in our simulations keeping the 
Keldysh parameter $\gamma$ constant, we find that the agreement between these three forms 
of harmonic spectra gets better. This 
observation is in agreement with the findings in Ref.~\cite{bandrauk}, because to keep 
$\gamma$ fixed, we scale the pulse duration by $\sim$$n^3$ and the peak 
laser field strength by $\sim$$1/n^4$. Although the energy of the initial state 
is also scaled by $\sim$$1/n^2$ and the pulse duration is the same in number of optical 
cycles, the ionization probability drops within a given 
$n$-series in Fig.~\ref{fig_hhg1d_iprob}. This suggests that the pulse is effectively 
getting weaker as we increase $n$ for fixed $\gamma$. We report only the 
dipole acceleration form $|a_z(\omega)|^2$ to refer to harmonic spectra, chiefly 
because it is this form that is directly proportional to the emitted power, 
{\it i.e.}, $S(\omega)=2\omega^4 |a_z(\omega)|^2 /(3\pi c^3)$. 

Because high-harmonic generation and ionization are competing processes in the 
physical regime we are interested in, it is useful to investigate the momentum 
distribution of the ionized part of the wave function to gain further insight 
into the HHG process. 
In order to evaluate the momentum distributions, we follow the same procedure 
outlined in Ref.~\cite{vrakking}. For sake of completeness, we briefly describe 
the method: In all simulations, the ionized part of the wave function is 
removed from the box every time step during the time propagation, 
in order to prevent unphysical reflections from the radial box edge. This is done 
by multiplying the wavefunction by a mask function $m(r)$ at every time step, 
where $m(r)$ spans 1/3 of the radial box at the box edge. We 
retrieve the removed part of the wave function by evaluating 
\begin{equation}\label{scheq3d:masked}
\Delta \psi_{l}(r,t')=[1-m(r)]\;\psi_{l}(r,t')
\end{equation}
at every time step, and Fourier transform it to get the momentum space wave 
function 
$\Delta \phi(p_{\rho},p_z,t')$, 
\begin{align}\label{scheq3d:pmap}
\Delta\phi(p_{\rho},p_z,t') = 2\;&\sum_{l}(-i)^{l}
\;Y_{l,m}(\theta,\varphi) \nonumber \\ 
	&\times\int_0^{\infty} j_{l}(pr) \Delta\psi_{l}(r,t') r^2 \;dr \;. 
\end{align}
Here the momentum $p=(p_{\rho}^2+p_z^2)^{1/2}$ is in cylindrical coordinates and 
$j_{l}(pr)$ are the spherical Bessel functions. We then time propagate 
$\Delta\phi(p_{\rho},p_z,t')$ to a later final time $t$ using the semi-classical 
propagator,
\begin{equation}\label{scheq3d:pprop}
\Delta \phi(p_{\rho},p_z,t) = \Delta \phi(p_{\rho},p_z,t')\;e^{-i S}
\end{equation}
where $S$ is the classical action. For the time-dependent laser field $F(t)$, 
action $S$ is calculated numerically by integrating $p_z^2$ along the laser 
polarization, 
\begin{eqnarray}\label{scheq3d:action}
S &=& \frac{1}{2}p_{\rho}^2 (t-t') + \frac{1}{2}\int_{t'}^t p_z^2 dt'' \\
p_z &=& \int_{t'}^t F(t'') dt'' 
\end{eqnarray}
We are assuming that the ionized electron is freely propagating in the classical 
laser field in the absence of the Coulomb field of its parent ion, and this method 
is numerically exact under this assumption. 

% 3d hhg spectra for 1s, 4s, and 8s
\subsection{Results and discussion} \label{sec:results_hhg3d}
\begin{figure}[ht!]
  \begin{center}
    \resizebox{150mm}{!}{\includegraphics{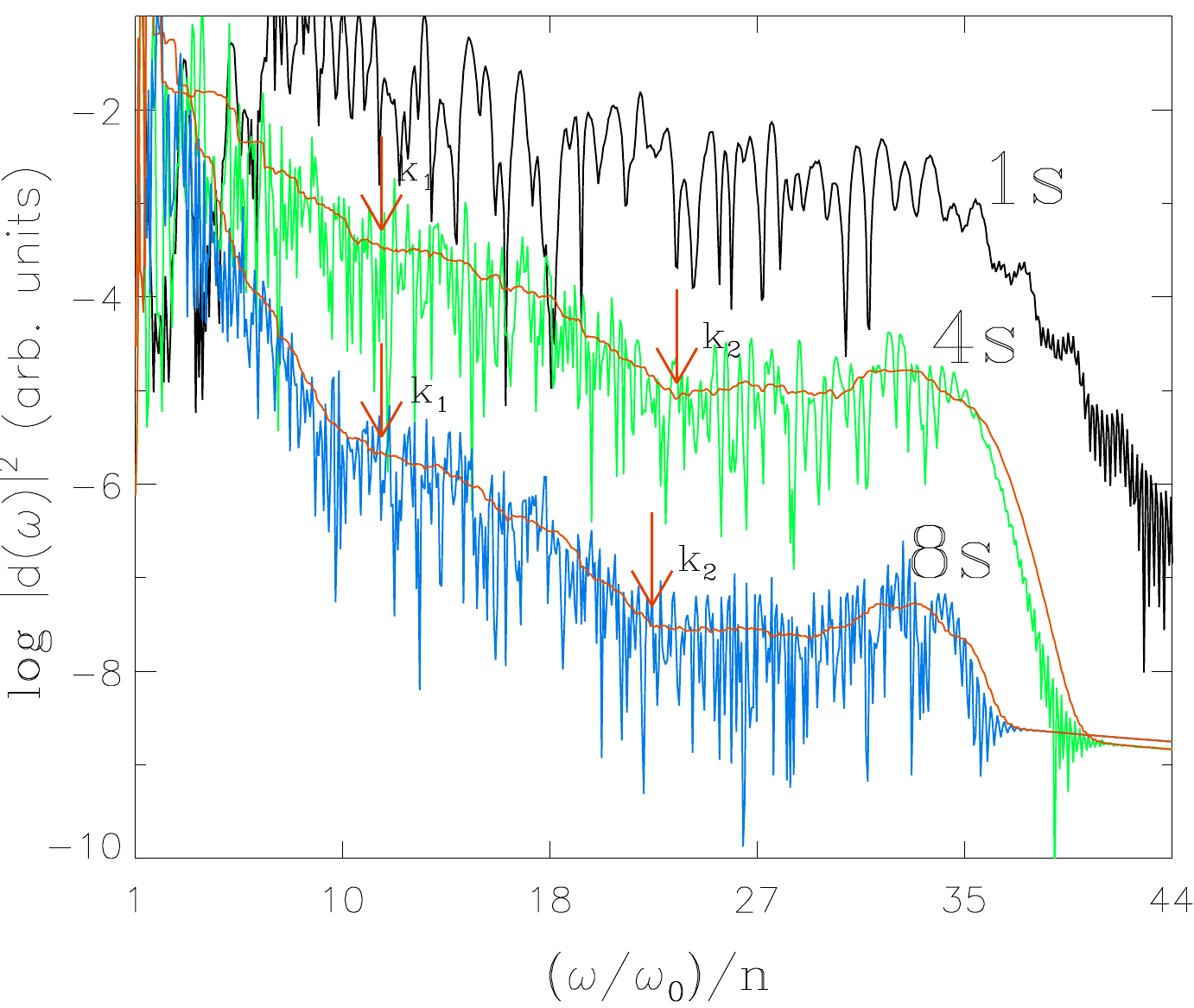}}
  \end{center}
  \caption{(Color online) Dipole acceleration from direct solution of the three-dimensional 
  time-dependent Schr\"odinger equation when the atom is initially prepared in 1s, 4s 
  and 8s states of H atom. The horizontal axis is the scaled harmonic order 
  $\widetilde{q}\equiv q/n=(\omega/\omega_0)/n$. There are three universal cut-off points 
  in the spectra: marked as $k_1$, $k_2$, and the usual $I_p+3.17 U_p$ limit. 
  The double plateau structure mirrors that of the 
  one-dimensional spectra from Fig.~\ref{fig_hhg1d}, with a universal secondary cut-off at 
  $\widetilde{q}=23.45$. The arrows marked as $k_1$ and $k_2$ are discussed in 
  the context of Fig.~\ref{fig_hhg3d_pmap2}. 
  }
  \label{fig_hhg3d}
\end{figure}

The double plateau structure we see in the one-dimensional spectra in 
Fig.~\ref{fig_hhg1d} can be also observed from our three-dimensional simulations. 
In Fig.~\ref{fig_hhg3d}, the squared dipole acceleration $|a(\omega)|^2$ is plotted 
for the initial states of 1s (black), 4s (green), and 8s (blue) of Hydrogen atom as 
a function of the scaled harmonic order $\omega/(\omega_0 n)\equiv \widetilde{q}$. 
In these calculations, we adhere to $\gamma=0.75$ as in the one-dimensional calculations, 
and start at $n=1$ with intensity $2\times10^{14}$ W/cm$^2$ and $\lambda=800$ nm. From 
this, we use the $n$ scaling discussed in Sec.~\ref{sec:results_hhg1d} to determine the 
laser parameters for higher $n$ states. Apart from the double plateau structure, there is 
decrease in the HHG yield with increasing $n$ in Fig.~\ref{fig_hhg3d}, similar to the 
one-dimensional case. Again, this suggests that although $\gamma$ is fixed for all three 
initial states in Fig.~\ref{fig_hhg3d}, the atom sinks deeper into the tunneling regime as 
$n$ is increased, similar to what we have seen in the one-dimensional case in 
Sec.~\ref{sec:results_hhg1d}. The main difference in Fig.~\ref{fig_hhg3d} is that the 
first plateau is not as flat as in the one-dimensional calculations, as often the case when 
comparing one- and three-dimensional HHG spectra. 

In order to clearly identify the first and the second cut-offs seen in Fig.~\ref{fig_hhg1d}, 
we have smoothened the 4s and 8s spectra by boxcar averaging to reveal their main structure 
(solid red curves) in Fig.~\ref{fig_hhg3d}. The usual scaled cut-off from the semiclassical 
three-step model is at $q_{\rm max}/n\simeq 35$ in all three spectra, and it is independent 
of $n$. A secondary cut-off emerges at the same scaled harmonic as in the one-dimensional case, 
which is labeled as $k_2$ in the 4s and the 8s spectra at $\widetilde{q}\simeq 23.45$. It is 
clear from Fig.~\ref{fig_hhg3d} that just as the usual cut-off at $q_{\rm max}/n$, $k_2$ 
is also universal beyond $n>4$. This secondary cut-off separates the two plateaus, first 
spanning lower frequencies below $k_2$, and the second spanning higher frequencies between 
$k_2$ and $q_{\rm max}/n$. 

The mechanism behind the formation of the secondary cut-off $k_2$ can be understood in 
terms of the ionization and the recombination steps of the semiclassical model. In the first 
step, the electron tunnels out of the initial $n$s state into the continuum, and has initially 
no kinetic energy. After excursion in the laser field, it recombines with its parent ion. In 
this last step, recombination occurs primarily back into the initial state. This is because the 
electron was liberated into the continuum with virtually no excess kinetic energy, and the 
electron wavepacket mainly retains its original character. When it returns to its parent ion 
to recombine, the recombination probability is highest for the bound state with which it overlaps 
the most. As a result, recombination into the same initial state is favored. This mechanism is 
associated with the usual cut-off since its position depends on the ionization potential: 
$q_{\rm max}=(I_p + 3.17 U_p)/\omega_0$. 

On the other hand, there is still probability that the electron can recombine to higher $n$ 
states. This would result in lower harmonics because less than $I_p$ needs to be converted 
to harmonics upon recombination. The cut-off for this mechanism would be achieved when the 
electron recombines with zero energy near the threshold ($n\rightarrow \infty$). Because 
the maximum kinetic energy a free electron can accumulate in the laser field is $3.17 U_p$, 
the lower harmonic plateau would cut off at $3.17 U_p$. For the laser parameters used in 
Fig.~\ref{fig_hhg3d}, this corresponds to the scaled harmonic $\widetilde{q}=23.45$, which is 
marked by the red arrows labeled as $k_2$ on the 4s and the 8s spectra. To reiterate, the 
second plateau with higher harmonics includes: 
\begin{enumerate}
\item trajectories which recombine to the initial state ($n_1 \rightarrow n_1$) after 
accumulating kinetic energy up to $3.17 U_p$,
\item trajectories which recombine to a higher but nearby $n$ state ($n_1 \rightarrow n_2$, 
where $n_2 > n_1$) that have acquired kinetic energy up to $3.17 U_p$, 
\item trajectories which recombine to much higher $n$ states ($n_1 \rightarrow n_2$, where 
$n_2 \gg n_1$) resulting in the cut-off at $\widetilde{q}=23.45$. 
\end{enumerate}

\begin{figure}[ht!]
  \begin{center}
    \resizebox{110mm}{!}{\includegraphics{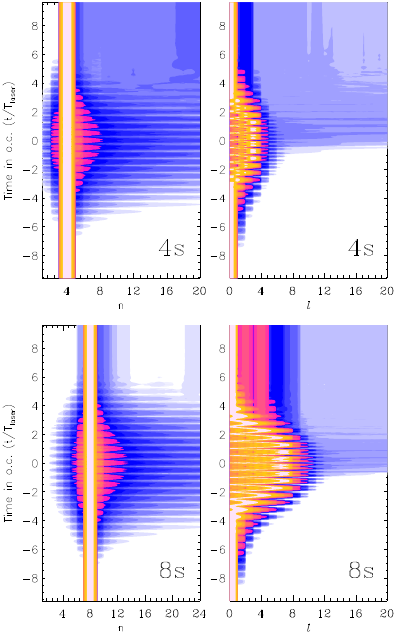}}
  \end{center}
  \caption{(Color online) $n$- and $l$-distributions for the probability to find the atom 
  in 4s and 8s states of H for the laser parameters used in Fig.~\ref{fig_hhg3d}. 
  All probabilities are plotted in $\log_2$ scale and the lowest contour in the 
  $n$-distributions for both states is at the 10$^{-10}$ level. The half-cycles of the 
  4-cycle laser pulse are clearly visible. 
  }
  \label{fig_hhg3d_nldist}
\end{figure}

The $n$- and $l$-distributions for the 4s and 8s states as a 
function of time can be seen in Fig.~\ref{fig_hhg3d_nldist}. Notice that the laser 
pulse is centered at $t=0$ o.c. and has 4 cycles at FWHM for both states. It is 
clear from the first column that the the atom mostly stays in the initial state and 
only a small fraction of the wavefunction contributes to the HHG process. 
To appreciate how small, we note that the highest contour is at unity, and lowest 
contour for both the 4s and the 8s states are at the $\sim$10$^{-10}$ level. At the end of 
the pulse, there is a small spread in $n$, which is skewed towards higher $n$ in 
both cases. This skew is expected since the energy separation between the adjacent $n$ 
manifolds drop as $\sim$$1/n^3$, and therefore it is easier to spread to the higher $n$ 
manifolds than to lower $n$. The small amplitude for this spread is a consequence of the 
fact that we are not in the $n$-mixing regime. In the second column, we see that the 
orbital angular momentum $l$ also spreads to higher $l$ within the initial $n$-manifold, 
and the small leakage to higher angular momenta at the end of the pulse is a consequence 
of the small probability for spreading to the higher $n$-manifolds. 

The second step of the harmonic generation process involving the free evolution of the 
electron in the laser field be understood on purely classical grounds. 
It was the classical arguments that led to the $3.17 U_p$ limit for the maximum 
kinetic energy attainable by a free electron. In the context of this paper, performing 
such classical simulations can yield no insight to how the excursion step of the HHG 
behaves under the scaling scheme we have employed so far. This is because the classical 
equations of motion perfectly scale under the transformations $r\rightarrow r n^2$, 
$t\rightarrow t n^3$, $\omega\rightarrow \omega/n^3$, and $E\rightarrow E/n^2$, 
where $r$ is distance and $t$ is time. 
On the other hand, it is the lack of this perfect scaling property of the Schr\"odinger 
equation that accounts for the differences between different initial $n$ states 
we have seen from our quantum simulations. One way to examine the excursion step by 
itself in our quantum simulations is to look at the momentum distribution of the part of 
the wavefunction that contributes to the HHG spectra. 

\begin{figure}[ht!]
  \begin{center}
    \resizebox{150mm}{!}{\includegraphics{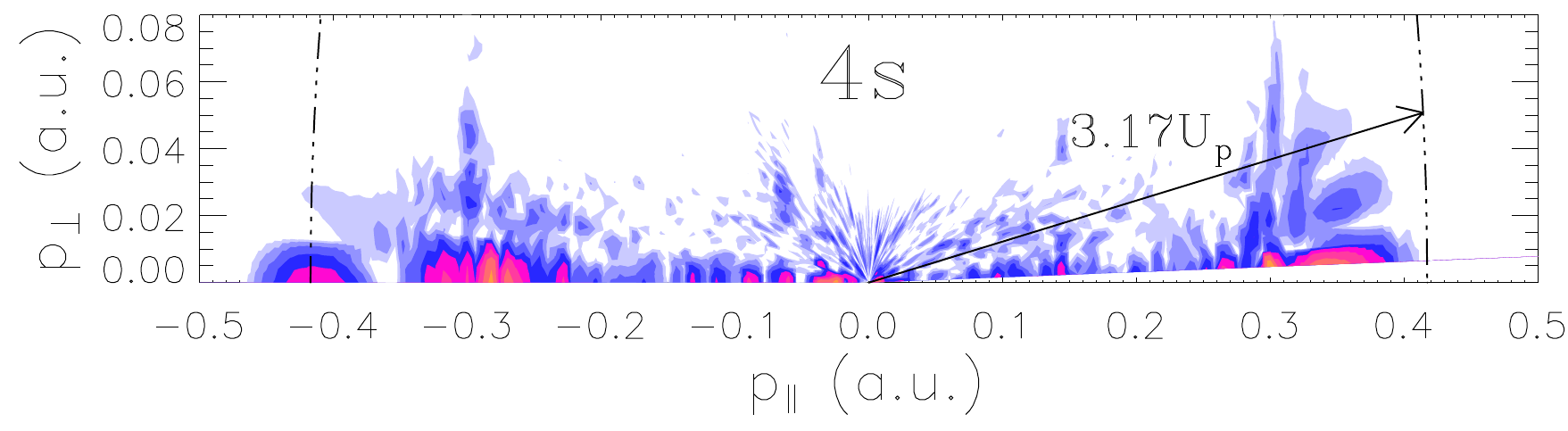}}
  \end{center}
  \caption{The momentum distribution for the ionized part of the wave function 
  integrated over time until after the laser pulse when the atom is initially 
  prepared in the 4s state. The total momentum $\sqrt{p^2_{||}+p^2_{\perp}}$ 
  corresponding to the maximum kinetic energy that can be attained by a free 
  electron in a laser field is marked by the dot-dashed semicircle and 
  labeled as $3.17 U_p$. This is the limit that determines the semiclassical 
  cut-off at $q_{\rm max}=I_p + 3.17 U_p$.  
  }
  \label{fig_hhg3d_pmap}
\end{figure}

To this end, we calculate the momentum map of the ionized part of the wavefunction when 
the atom is initially prepared in the 4s state. The reason we look at the ionized part 
of the wavefunction is because harmonic generation and ionization are competing processes. 
Therefore one would expect that they should mirror each other in their behavior. 
Fig.~\ref{fig_hhg3d_pmap} shows this momentum distribution obtained by Fourier transforming 
the ionized part of the wavefunction, which is accumulated over time until after the 
laser pulse (see Eq.~\eqref{scheq3d:masked} onward). Since the problem has cylindrical 
symmetry, the horizontal axis is labeled $p_{||}$ to refer to the momentum component 
parallel to the laser polarization direction (same as $p_z$). The vertical axis 
$p_{\perp}$ is the perpendicular component. We have also labeled the $3.17 U_p$ limit 
for the maximum kinetic attainable, which is along the dot-dashed semicircle. As expected, 
the total momentum of the escaped electrons cut off at $3.17 U_p$, and the components 
which would have contributed to the two different plateaus in Fig.~\ref{fig_hhg3d}  
are visible close to the laser polarization direction. 

\begin{figure}[h!tb]
	\begin{center}$
		\begin{array}{cc}
		   \resizebox{74mm}{!}{\includegraphics{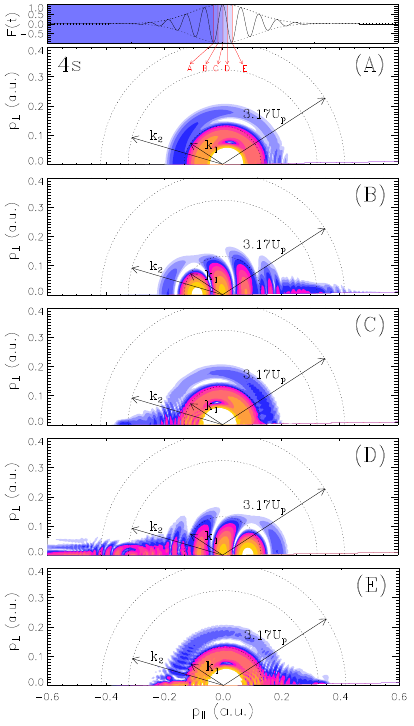}} & 
       \resizebox{84mm}{!}{\includegraphics{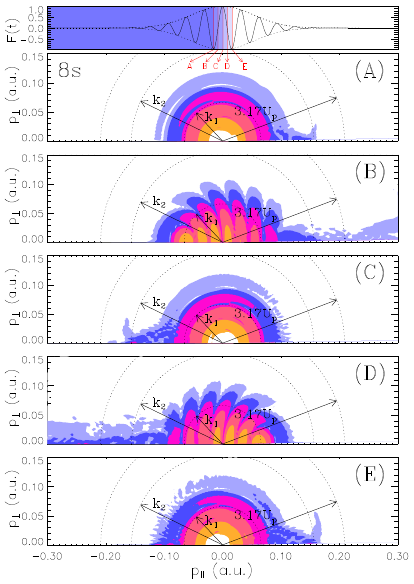}}  
         \end{array}$
     \end{center}
  \caption{(Color Online) Momentum distributions in the region $r>1/\sqrt{F}$ 
  for the 4s (left column) and 8s (right column) states at five instances during 
  the laser cycle at the peak of the pulse (indicated on top). The region 
  $r>1/\sqrt{F}$ is beyond the peak of the Coulomb potential depressed by 
  the strong laser field. 
  }
  \label{fig_hhg3d_pmap2}
\end{figure}

We also look at the momentum map of the wavefunction inside our numerical box that 
falls beyond the peak of the depressed Coulomb potential at $r=1/\sqrt{F}$. 
Part of the wavefunction in the 
region $r < 1/\sqrt{F}$ is removed by multiplying it with a smooth mask 
function before the Fourier transformation step described in Sec.~\ref{sec:theory_hhg3d}. 
The results when the atom is initially in the 4s and 8s states are seen in 
Fig.~\ref{fig_hhg3d_pmap2} at five instances during the laser cycle at 
the peak of the pulse (labeled A, B, C, D, and E). We have also labeled three semicircles 
corresponding to three momenta $\sqrt{p^2_{||} + p^2_{\perp}}$ of interest: 
\begin{enumerate}
\item the $3.17 U_p$ limit, also seen in Fig.~\ref{fig_hhg3d_pmap}, 
\item  $k_1$ corresponding to the kinetic energy $U_p$, 
\item $k_2$ corresponding to the kinetic energy necessary to emit the 
harmonic $\widetilde{q}=23.45$ at the secondary cut-off in Fig.~\ref{fig_hhg3d}, 
{\it if the electron recombines into its initial 4s or 8s state upon rescattering}. 
\end{enumerate}
The amplitude inside the $k_1$ semicircle contributes to only very low harmonics, 
below the scaled harmonic labeled as $k_1$ in Fig.~\ref{fig_hhg3d}. This part of the 
spectra is not suitable for the semiclassical three step description of HHG. The annular region 
between the semicircles $k_1$ and the $k_2$ contributes to the first low harmonic 
plateau in Fig.~\ref{fig_hhg3d}. Finally, the region between $k_2$ and the semiclassical 
$3.17 U_p$ limit contributes to the less intense second plateau. The distinction 
between the lower harmonics from the inner $k_1$ semicircle and the higher harmonics from 
the annular region between $k_1$ and $k_2$ is manifested most clearly in the 4s column, 
as longer and shorter wavelengths in the momentum maps inside these regions. Expectedly, 
both momentum maps for the 4s and the 8s initial states show the same structures, the essential 
difference being the number of nodes in the momentum space wave functions which scales as $n^2$. 
Incidentally, a rescattering event is visible on the laser polarization axis at $k_2$ in panel 
D of the 4s column, giving rise to kinetic energy beyond the $3.17 U_p$ limit on the left. 

\section{Conclusions}
We have presented results from one- and three-dimensional time-dependent quantum 
calculations for higher-order harmonic generation from excited states of H atom 
for a fixed Keldysh parameter $\gamma$. Starting from the ground state, we chose 
laser intensity and frequency such that we are in the tunneling regime and ionization 
probability is well below one per cent. We then scale the intensity by $1/n^8$ and the 
frequency by $1/n^3$ to keep $\gamma$ fixed as we increase the principal quantum 
number $n$ of the initial state of the atom. Because $\gamma$ is fixed, the common 
wisdom is that the dynamical regime which determine the essential physics should 
stay unchanged in the HHG process as we go up in $n$ of the initial state. 
Our one-dimensional calculations demonstrate that this is indeed the case, and 
although the emitted power (HHG yield) drops as we climb up in $n$, the resulting 
harmonic spectra display same {\it universal} features beyond $n$$\sim$10. The most 
distinguished feature that develops when the atom is initially prepared in a Rydberg state 
is the emergence of a secondary plateau below the semiclassical cut-off $q_{\rm max}$
in the HHG plateau. This secondary cut-off splits the harmonic plateau into two regions: 
one spanning low harmonics and terminating with a secondary cut-off, and a second 
plateau with lower yield and higher harmonics terminating at the usual semiclassical 
cut-off at $q_{\rm max}$. 

We have also found that the positions of these cut-off harmonics also scale 
as $1/n$, and introduced the concept of ``scaled harmonic order", 
$\widetilde{q}=\omega/(\omega_0 n)$. When plotted as a function of $\widetilde{q}$, 
the harmonic spectra appear universal and, except for the overall yields, the spectra 
for high $n$ look essentially identical. 

We then carried out fully three-dimensional calculations for three of the $n$ states in the 
lowest $n$-group in the one-dimensional calculations to gain further insight into 
the scaling properties we have seen in the one-dimensional calculations. This also 
serves to investigate possible effects of having angular momentum. We found the same features 
as in the one-dimensional spectra, except that the yield from the first plateau is 
skewed towards lower harmonics. We associate this with spreading to higher $n$ states 
during the tunnel ionization and recombination steps by analyzing the 
$n$- and $l$-distributions of the atom after the laser pulse. Momentum distributions 
of the ionized electrons and the wave function beyond the peak of the depressed 
Coulomb potential at $r=1/\sqrt{F}$ show features which we can relate to the universal features 
we see in the HHG spectra at high $n$. We identify the first plateau in this universal 
HHG spectrum with features in momentum space between two values of momentum: (1) the momentum 
corresponding to kinetic energy $U_p$, and (2) the momentum corresponding to 
kinetic energy if the electron emits the secondary cut-off harmonic 
upon recombining to its initial state. The latter case also occurs when the electron 
recombines to a much higher Rydberg state than the one it tunnels out after 
accumulating maximum possible kinetic energy of $3.17 U_p$ during its excursion in 
the laser field.

\section{Acknowledgments}
IY, EAB and ZA was supported by BAPKO of Marmara University. ZA would like to thank to 
the National Energy Research Scientific Computing Center (NERSC) in Oakland, CA. 
TT was supported by the Office of Basic Energy Sciences, US Department of Energy, and by 
the National Science Foundation Grant No. PHY-1212482.

\onecolumngrid

% 1d quantum calculations
\newpage
\framebox[1.4\width]{\Large{\color{red}{Fig. 01}}} 
\vspace{4cm}

\newpage
\framebox[1.4\width]{\Large{\color{red}{Fig. 02}}} 
\vspace{4cm}

\newpage
\framebox[1.4\width]{\Large{\color{red}{Fig. 03}}} 
\vspace{4cm}

% 3d quantum calculations
\newpage
\framebox[1.4\width]{\Large{\color{red}{Fig. 04}}} 
\vspace{4cm}

\newpage
\framebox[1.4\width]{\Large{\color{red}{Fig. 05}}} 
\vspace{2cm}

\newpage
\framebox[1.4\width]{\Large{\color{red}{Fig. 06}}} 
\vspace{6cm}

\newpage
\framebox[1.4\width]{\Large{\color{red}{Fig. 07}}} 
\vspace{1cm}

\end{document}